\documentclass{article}
\usepackage{aaai17}
\usepackage{url}
\usepackage{amsfonts}
\usepackage{amssymb}
\usepackage{amsmath}
\usepackage{graphicx}
\usepackage{longtable}
\newtheorem{definition}{Definition}

\nocopyright

\title{Privacy of Existence of Secrets:\\
Introducing Steganographic DCOPs\\
and Revisiting DCOP Frameworks}
\author{
Viorel D. Silaghi and Marius C. Silaghi and Ren\'e Mandiau\\
Florida Institute of Technology\\
University of Valenciennes\\
\url{vsilaghi2018@my.fit.edu}, \url{msilaghi@fit.edu}, \url{rene.mandiau@univ-valenciennes.fr}
}

\begin{document}
\maketitle{}

\begin{abstract}
Here we identify a type of privacy concern in Distributed Constraint Optimization (DCOPs) not previously addressed in literature, despite its importance and impact on the application field: {\em the privacy of existence of secrets}.
Science only starts where metrics and assumptions are clearly defined.
The area of Distributed Constraint Optimization has emerged at the intersection of the multi-agent system community and constraint programming. For the multi-agent community,
the constraint optimization problems are an elegant way to express many of
the problems occurring in trading and distributed robotics. For the theoretical constraint programming community the DCOPs are a natural extension of their main object of study, the constraint satisfaction problem.
As such, the understanding of the DCOP framework has been refined with the needs of the two communities, but sometimes without spelling the new assumptions formally and therefore making it difficult to compare techniques. Here we give a direction to the efforts for structuring concepts in this area.
\end{abstract}

\section{Introduction}

We identify the {\em ``privacy of existence of secrets''} as a not yet formally explored issue in the area of the Distributed Constraint Optimization problems (DCOPs).
In order to scientifically draw conclusions about a concept or procedure, it is essential to formally define the involved concepts, metrics, and assumptions. Consequently, we also revisit some common concepts to suggest a direction for the possible structuring of DCOP frameworks.

The distributed constraint optimization has emerged at the confluence of research in Multi-Agent Systems and research in Constraint Programming.
The multi-agent community has found in the DCOP framework an elegant way to specify requirements existing in the area of cooperating robot teams, negotiation, and trading.
For the Constraint Programming community, the DCOP is a natural extension of a framework 
that is at the center of their field, the constraint satisfaction problem (CSP).
The reasons and the expectations from the new DCOP framework are slightly different for
the researchers in the involved areas, differences that for various reasons are sometimes not spelled formally inside the DCOP definitions, making comparison and reference difficult and error-prone for new researchers in this booming area.

The Constraint Programming community has developed an impressive body of knowledge and set of techniques  focused on improving the efficiency of problems represented as constraint satisfaction problems (CSP) and for various extensions of such problems including the egalitarian Fuzzy CSPs (FCSPs), Probabilistic CSPs (PCSPs), dynamic CSPs (DCSPs), the utilitarian Weighted CSPs (WCSPs), and the general Valued CSPs (VCSPs)~\cite{schiex1995valued} and Semiring based CSPs~\cite{Bistarelli1999}.
The Constraint Programming community is careful in formally defining each extension to CSPs and clearly stating each framework by specifying formally and exactly the inputs and
unambiguously defining outputs and evaluation metrics. This allows for work in the Constraint Programming field to be largely unambiguously comparable, and technology transfer between extensions to be possible without need of repeating work. However, most research in that area concerns efficiency while privacy and secrecy of constraints are generally not part of the CSP community criteria.
An early set of extensions of the CSP family to the multi-agent world addressed parallelism~\cite{kasif1990parallel}, and continued to be focused solely on efficiency, as
constraint were considered shared between agents.

The seminal work 
in~\cite{yokoo1992distributed} brought the concept of a distribution of
CSPs where privacy was mentioned as a motivation, besides natural distributions of the
constraints and control of variables to agents, as a reason to not attempt sharing constraints except on a need basis during
a distributed backtracking effort.
While this is also seen as an extension of the CSP family, the efficiency was no longer a sufficient criteria and direct comparison with traditional centralized constraints was
seen as irrelevant, despite a lack of formal specification of the impediments to additional constraint sharing in terms of either technical cost or quantified privacy constraints.
This lack of specification gave birth to a flurry of works exploiting various amounts of constraint information without quantification and with efficiency gains whose trade-off was
not always clear~\cite{silaghi2000distributed} in terms of the privacy or in terms of the technical cost of constraint data centralization.

In a concern to make the field coherent and easily maintain scientific soundness,
the constraint community recommended at the CP2000 distributed constraint satisfaction workshop panel
to name the new family of extensions by appending {\em distributed} in front of each CSP extension, yielding:
\begin{itemize}
\item distributed constraint satisfaction (DisCSP) 
\item distributed fuzzy constraint satisfaction (DisFCSP) 
\item distributed probabilistic constraint satisfaction (DisPCSP) 
\item distributed dynamic constraint satisfaction (DisDCSP or DisDyCSP) 
\item distributed weighted constraint satisfaction (DisWCSP) 
\item distributed valued constraint satisfaction (DisVCSP) 
\item distributed semiring-based constraint satisfaction (DisSCSP) 
\end{itemize}

Works had started to confusingly use the acronym DCSP for distributed CSPs, conflicting with the acronym for dynamic CSPs.
It was recommended for the new frameworks to use the prefix "Dis" rather than "D" as a way
to differentiate from dynamic CSPs which were intensively studied at the time. It was suggested that the latter be referred as DyCSP to further reduce confusion. Later, additional types of problem emerged, highlighting the wisdom of the suggestion, namely:
\begin{itemize}
\item dynamic distributed constraint satisfaction (DyDisCSP)
\item dynamic distributed dynamic constraint satisfaction (DyDisDyCSP)
\end{itemize}
where besides a dynamism in constraint emergence there also exist a dynamism in agent involvement (i.e., churning) or in constraint redistribution~\cite{silaghi2001generalized,silaghi2002openness}.

However, while privacy was already a stated motivation for distributed CSPs, (and by extension assumed present in all these frameworks), it remained unspecified formally,
perpetuating the difficulty of correct comparison and evaluation of proposed techniques.

Meanwhile, the multi-agent community adopted the new direction and developed techniques coining the name ``distributed constraint optimization'' (DCOP) for a framework of 
distributed weighted constraints. For the early DCOPs definitions, privacy was also not formally stated, and the names DCOP and DisWCSP coexisted for a while until the community
largely adopted the DCOP acronym. Extensions in this community are built on the DCOP concept,
such as in the case of Leximin DCOPs~\cite{matsui2018leximin}.

Further research noticed the negative effects from lack of formal definitions and quantification of privacy 
criteria in the distributed CSP family. Attempts to fix the problem either 
redefined the framework~\cite{silaghi2005desk} without changing its name/acronym (in an effort
to redeem or adopt earlier work that stressed privacy), or alternatively proposed new names and 
acronyms. New proposed names/acronyms appeared in the case of distributed Privacy CSPs (DisPrivCSPs) or distributed privacy DCOPs (DisPrivCOPs)~\cite{doshi2008distributed} and Utilitarian CSPs (UCSPs) or Utilitarian DCOPs (UDCOP)~\cite{savaux2016discsps,savaux2017utilitarian}.
Work has also addressed the classification of the nature of privacy concerns, as in~\cite{leaute2013protecting}.

In this work, privacy is defined as a value:
\begin{definition}
Privacy is the utility that agents draw from conserving the secrecy of their personal information.
\end{definition} 

In the Background section,
we review the original and main formal definitions of distributed constraint satisfaction (DisCSP) and distributed constraint optimization problems (DCOPs/DisWCSPs), as well as the main informal privacy classifications in literature.
We also introduce the concept of steganography, which is more intensively studied in the area of Cryptology.
In the Steganographic DCOP section
we then present motivation and define formally the Steganographic DCOPs.
Further in the Frameworks Disambiguation section
we propose a direction for the structure of the nomenclature and definitions for various concepts in the DCOP areas.
We conclude with a summary of the contributions.

\section{Background}\label{sec:background}

We start by reviewing the main constraint optimization frameworks as emerging from constraint satisfaction. Later we review the various classifications of privacy found in literature, and the definitions we used for privacy and steganography.

\subsection{Formal Distributed Constraint Optimization Frameworks}
Several distributed CSPs are based on the CSP framework:

\begin{definition}[CSP~\cite{yokoo1998distributed}]
A CSP consists of $n$ variables $x_1, x_2, ..., x_n$, whose values are taken from finite, discrete domains
$D_1, D_2, ..., D_n$ respectively, and a set of constraints on their values. A constraint is defined by a
predicate. That is, the constraint $p_k(x_{k1},...,x_{kj})$ is a predicate that is defined on the Cartesian 
product $D_{k1}\times ...\times D_{kj}$. This predicate is true iff the value assignment of these variables satisfies this
constraint. Solving a CSP is equivalent to finding an assignment of values to all variables such that
all constraints are satisfied.
\end{definition}

The first distributed CSP definition from~\cite{yokoo1992distributed,yokoo1998distributed} is:

\begin{definition}[DCSP~\cite{yokoo1998distributed}]
\label{def:DCSP}

There exist $m$ agents: $1, 2, ...,m$. Each variable $x_j$ belongs to one agent i (this relation is represented as
$belongs(x_j, i))$. Constraints are also distributed among agents. The fact that an agent $l$ knows a
constraint predicate $p_k$ is represented as $known(p_k,l)$.

A Distributed CSP is solved iff the following conditions are satisfied:

$\forall i, \forall x_j$ where $belongs(x_j, i)$, the value of $x_j$ is assigned to {some value} $d_j$, and $\forall l, \forall p_k$ where $known(p_k, l)$, $p_k$ is true under the assignment $x_j = d_j$.
\end{definition}

As observed, even as~\cite{yokoo1998distributed} has a section dedicated to privacy motivations, the formal definition does not specify a way to quantify privacy requirements and costs. Further, the semantic of variable ownership specified with the $belongs$ predicate is not further formally defined
except as implied by the fact that in the proposed algorithms, new assignments for a variable are only proposed by the agents to which the variable $belongs$.

While this definition assumes that the number of agents equals the number of variables, with each agent keeping secret the domain of exactly one variable, other versions studied the case of unequal numbers of variables and agents with no secret domains but with secret constraints~\cite{silaghi2000asynchronous}.

\begin{definition}[DisCSP~\cite{silaghi2005asynchronous}]\label{def:DisCSP}
Distributed constraint satisfaction problems (DisCSPs) is defined by:
\begin{itemize}
\item
a set of $n$ variables $X = \{x_1, ..., x_n\}$,
\item
a set of $n$ domains, $D = \{d_1, ..., d_n\}$, for the variables,
\item
a set of $t$ constraints, $C = \{c_1 = (x_i
, x_j, ...), ...,c_t\}$, each of which is a subset of
the set of variables, linked with a relation, and
\item
a set of $t$ relations, $R = \{r_1, ..., r_t\}$. $r_i$ gives the allowed value combinations for
the corresponding constraint $c_i$.
\item 
a set of $m$ independent but communicating agents $A = \{A_0, .., A_m\}$
\item 
an ownership mapping $M : X \cup C \rightarrow {\cal P}(A)$ that assigns each variable or constraint to the subset of agents that own it.
 ${\cal P}(A)$ is a common
notation for the set of subsets of $A$.
\end{itemize}

A solution to a CSP is an assignment of values from the corresponding domains
to each variable such that for all constraints, the combination of assigned values
is allowed by the corresponding relation.
\end{definition}

The  Distributed Constraint Optimization Framework was introduced in~\cite{modi2002asynchronous}.

\begin{definition}[DCOP~\cite{modi2002asynchronous}]
\label{def:DCOP}
A Distributed Constraint Optimization Problem (DCOP) consists of $n$ variables $V = \{x_1,x_2,...x_n\}$, each assigned to an agent, where the values of the variables are taken from finite, discrete domains $D_1, D_2,...D_n$, respectively. Only the agent who is assigned a variable has control of its value and knowledge of its domain.

The goal is to choose values for variables such that an objective function is minimized or maximized. The objective function described is addition over costs, but  can be any associative, commutative,
monotonic aggregation operator defined over a totally ordered set of valuations, with
minimum and maximum element (described by Schiex, Fargier and Verfaillie as Valued CSPs~\cite{schiex1995valued}).
\end{definition}

The definition in~\cite{modi2002asynchronous} makes a formal link to the Weighted CSP instance of the Valued CSP framework while claiming generality in terms of application to remaining CSP extensions.

However, while further stressing the control of values, it limits input secrecy to domains and does not further quantify effects of privacy loss or penalties for discussions by agents about values that they do not ``control''.

Despite the framework not quantifying privacy, measurements of privacy loss based on qualitative comparisons were made in~\cite{silaghi2002comparison}, and based on using assumptions from information theory, outside the standard DCOP definitions, in~\cite{freuder2001privacy,franzin2002multi,wallace2004using,maheswaran2006privacy} and in ~\cite{greenstadt2006analysis,faltings2008privacy}.

An effort to redefine DisCSPs to formalize privacy requirements is in~\cite{silaghi2005desk}:

\begin{definition}[DisCSP~\cite{silaghi2005desk}]
A [Cryptographic] Distributed CSP (DisCSP) is defined by six sets $(A, X, D, C, I, O)$ and an algebraic structure $F$.
$A=\{A_1, ..., A_n\}$ is a set of agents. $X$, $D$, and the solution
are defined like for CSPs.

$I=\{I_1, ..., I_n\}$ is a set of secret inputs. $I_i$
is a tuple of $\alpha_i$
secret inputs (defined on $F$) from the agent $A_i$. Each input
$I_i$ belongs to $F^{\alpha_i}$.

Like for CSPs, $C$ is a set of constraints. There may exist
a public constraint in C, $\phi_0$, defined by a predicate $\phi_0(\varepsilon)$
on tuples of assignments $\varepsilon$, known to everybody. However,
each constraint $\phi_i,i>0$, in $C$ is defined as a set of known
predicates $\phi_i(\varepsilon,I)$ over the secret inputs $I$, and the tuples
$\varepsilon$ of assignments to all the variables in a set of variables $X_i$,
$X_i \subseteq X$.

$O=\{o_1, ..., o_n\}$ is the set of outputs to the different agents.
Let $m$ be the number of variables. 
$o_i : D_1\times ...\times D_m \rightarrow F^{\omega_i}$
is a function receiving as parameter a solution and returning
$\omega_i$ secret outputs (from $F$) that will be revealed only to the
agent $A_i$.
\end{definition}

This definition uses the same name and acronym as previously encountered definitions of distributed CSPs, but formally specifies that solutions may only be revealed to specific agents (introducing a concept now referred in literature as privacy of decision~\cite{leaute2013protecting}). The formulation requires that there should not be any kind of privacy loss, being mainly appropriate for cryptographic techniques. 

The extension of such privacy formalization to DCOPs is reported in~\cite{silaghi2005using}:

\begin{definition}[DisWCSP~\cite{silaghi2005using}]\label{def:DisWCSP}
A distributed constraint satisfaction problem (DisWCSP) is defined by six
sets $(A, X, D, C, I, O)$, an arithmetic structure $F$, and a set of acceptable solution qualities $B$, that can be often represented as an interval $[B_1, B_2]$. $A=\{A_1, ..., A_n\}$ is a
set of agents. $X = \{x_1, ..., x_m\}$ is a set of variables and $D = \{D_1, ..., D_m\}$ is a set of finite domains such that $x_i$ can take values only from 
$D_i = \{v^i_1,..., v^i_{d_i}\}$. 

An assignment
is a pair $\langle x_i, v^i_k\rangle$
meaning that the variable $x_i$ is assigned the value $v^i_k$. 

A tuple is an
ordered set. $I=\{I_1, ..., I_n\}$ is a set of secret inputs. $I_i$
is a tuple of $\alpha_i$ secret inputs
(defined on a set $F$) from the agent $A_i$. Each input $I_i$ belongs to 
$F^{\alpha_i}$. 

$C = \{\phi_0, ..., \phi_c\}$
is a set of constraints. A constraint $\phi_i$ weights the legality of each combination of
assignments to the variables of an ordered subset $X_i$ of the variables in $X$, $X_i \subseteq X$. 
$\phi_0$ is a public constraint defined by a function $\phi_0(\varepsilon)$ on tuples of assignments $\varepsilon$,
known to everybody. Each constraint $\phi_i,i>0$, in $C$ is defined as a known function
$\phi_i(\varepsilon,I)$ over the secret inputs $I$, and the tuples $\varepsilon$ of assignments to all the variables
in a set of variables $X_i$, $X_i \subseteq X$. 

The projection of a tuple $\varepsilon$ of assignments over a
tuple of variables $X_i$ is denoted $\varepsilon_{\mid X_i}$. 
A solution is:
$$\varepsilon^*= argmin_{\varepsilon\in D_1\times ... \times D_n} \sum_{i=1}^{c} \phi_i(\varepsilon_{\mid X_i}),$$ 
if
$\sum^c_{i=1} \phi_i(\varepsilon^*_{\mid X_i}) \in [B_1...B_2]$. 

$O=\{o_1, ..., o_n\}$ is the set of outputs to the different agents.
$o_i : I \times D_1 \times ... \times D_m \rightarrow F^{\omega_i}$
is a function receiving as parameter the inputs and
a solution, and returning $\omega_i$ secret outputs (from $F$) that will be revealed only to the
agent $A_i$. The problem is to generate $O$.
\end{definition}

The above DCOP framework was designed to be addressed with cryptographic protocols and allows for chaining multiple DCOPs for solving complex problems such as Vickrey Auctions~\cite{silaghi2005using}. It does not 
support non-cryptographic privacy aware solvers as it does not model
costs for partial privacy leaks,

\paragraph{Frameworks for DCOP With Privacy for Non-cryptographic Solvers}
The quantification of privacy in distributed CSPs is introduced in~\cite{savaux2016discsps}, to enable the evaluation of non-cryptographic solvers:

\begin{definition}[UDCSP]\label{def:UDCSP}
A UDisCSP is formally defined as a tuple
$\langle A, V, D, C, U, R\rangle$ where:
\begin{itemize}
\item $A = \langle A_1, ... , A_n\rangle$ is a vector of $n$ agents
\item $V = \langle x_1, ... , x_n\rangle$ is a vector of $n$ variables. Each agent $A_i$ controls the variable $x_i$.
\item $D = \langle D_1, ... , D_n\rangle$ where $D_i$ is the domain for the variable $x_i$, known only to $A_i$, and 
a subset of $\{1, ... , d\}$.
\item $C = \{C_1, ... , C_m\}$ is a set of interagent constraints.
\item $U=\{u_{1,1},...,u_{n,d}\}$ is a matrix of privacy costs where $u_{i,j}$ is
the cost of agent $A_i$ for revealing whether $j \in D_i$.
\item $R = \langle r_1, ... , r_n\rangle$ is a vector of rewards, where $r_i$
is the
reward agent $A_i$ receives if an agreement is found.
\end{itemize}
An agreement as a set of assignments for all
the variables with values from their domain, such that all the
constraints are satisfied.

The state of agent $A_i$ includes the subset of $D_i$ that it has
revealed, as well as the achievement of an agreement. The
problem is to define a set of communication actions and a
policy for each agent such that their utility is maximized.
\end{definition}

Another framework that allows for privacy loss but quantifies it was proposed in~\cite{doshi2008distributed}. It is defined formalizing the concept of revelation:

\begin{definition}[Revelation]
Given a set of Boolean (propositional) secrets $S$ and a set of agents $A$, 
a possible revelation
$R(A, S)$ is a function $R(A, S): A\times S \rightarrow [0, 1]$ which maps
each peer agent and a secret to the probability learned by
that agent about the secret.
\end{definition}

\begin{definition}[DPCOP~\cite{doshi2008distributed}]\label{def:DPCOP}
A (minimization) Distributed Private Constraint Optimization Problem (DPCOP) is defined by a tuple $(A, X, D, C, P, U)$. $A$ is a set of agents
$\{A_1, ..., A_K\}$. $X$ is a set of variables $\{x_1, ..., x_n\}$, and $D$
is a set of domains $\{D_1, ..., D_n\}$ such that each variable
$x_i$ may take values only from the domain $D_i$. The variables are subject to a set $C$ of sets of weighted constraints
$\{C_0, C_1, ..., C_K\}$, where $C_i = \{\phi_i^1 , ..., \phi_i^{c_i}\}$ holds the secret weighted constraints of agent $A_i$, and $C_0$ holds the
public constraints. Each weighted constraint is defined as a
function $\phi_i : X_i \rightarrow {\mathbb R}_+$ where $X_i \subseteq X$. The value of such
a function in an input point is called {\em constraint entry}, and
each $C_i$ can be seen as a set $C_i$ of such constraint entries. $P$
is a set of privacy loss cost functions $\{P_1, ..., P_K\}$, one for
each agent. $P_i$ defines the cost inflicted to $A_i$ by each revelation $r$ of its secrets, i.e., $P_i(r) : R(A, C_i) \rightarrow {\mathbb R}_+$. $U$ is a
set $\{U_1, ..., U_K\}$. $U_i$ is the reward received by $A_i$ if a solution
is found (used for deciding to abandon the search).

A solution is an agreement between agents in $A$ on a tuple $\epsilon^*$ of assignments of values to variables that minimizes the total cost:
$$\epsilon^∗ = argmin_\epsilon\sum_i\left(\sum_j \phi_j(\epsilon)\right)+P_i\left(\prod_i(\epsilon)\right)$$
where $\prod_i(\epsilon)$ is the revelation $R(A,C_i)$ during the process
leading to the agreement on the assignments $\epsilon$.

\end{definition}

This definition quantifies privacy and drops the concept of control by agents on variables, which we describe as unclear in previous definitions. A variant related to UDCSP is described and evaluated in~\cite{savaux2017utilitarian}.

\subsection{Privacy Classifications}

There are several efforts to classify privacy concepts in distributed constraint optimization~\cite{leaute2013protecting}.

\begin{itemize}

\item {\bf Domain Privacy As Existence of Values}
The explanation in~\cite{yokoo1998distributed} concerning the first distributed CSP definition can be interpreted as describing a privacy concerning the very existence of values in domains. This goes to the point where the constraints of other agents cannot be defined extensionally due to the impossibility
to enumerate these domains (e.g., impossibility to enumerate all meeting places that an agent controlling the corresponding variable may think about).
This kind of domain privacy was highlighted in~\cite{brito2003distributed}.

\item {\bf Domain Privacy As Unary Constraint}
The first types of privacy mentioned in~\cite{yokoo1998distributed} are domain and constraint privacy.
Domain privacy is frequently understood as a unary constraint on the values of a variable~\cite{yokoo1998distributed}.
This interpretation is assumed in many subsequent works, where agents' constraints are assumed to be extensionally defined using mutual knowledge of the values in domains
of each others' variables~\cite{silaghi2000asynchronous,modi2002asynchronous}.

\item {\bf Constraint Privacy}

Constraint privacy is linked to the $knows$ predicate in
the original distributed CSP definition~\cite{yokoo1998distributed}.
Later work propose to drop the variable controlling assumptions
in exchange for focusing on constraint privacy~\cite{silaghi2000distributed,silaghi2000asynchronous}.
Intermediary frameworks allowing for some constraint privacy by distribution in order to maintain the notion of control of variables have also been proposed by splitting binary constraints into private halves, but it is unclear if the under-defined control of variables offers a tradeoff for the reduced constraint privacy.

\item {\bf Privacy of Existence of Solution}

The fact that the problem in not solvable may in itself be a secret and impact
on revelation of information. This privacy was addressed first in~\cite{silaghi2005hiding}, based on stochastically discarding solutions even in  problems found solvable by systematic solvers, to hide solution existence.

\item {\bf Privacy of Decision}
The decision, as to what value is assigned to an individual variable in the final solution, only needs to be revealed to a predefined set of agents, and this was  introduced in~\cite{cryptoeprint:2004:333,silaghi2005desk}.

\item {\bf Privacy of Assignments}

The individual values being tried during search by an agent can be secret and certainly reveal secrets about unary constraint of the proposer. This privacy is arguably at least partially achieved by the original ABT algorithm in~\cite{yokoo1992distributed} where assignments are only announced to relevant agents.
A twist where values are only revealed by relation was sometimes also claimed to increase this privacy~\cite{silaghi2000asynchronous,brito2003distributed}.
It is unclear whether cryptographic algorithms offer this privacy or it is simply 
irrelevant for them, as no proposal is made in them~\cite{silaghi2004meeting,silaghi2004distributed}.

\item {\bf Privacy of Algorithms}

The privacy of algorithm is typically defined as a privacy with respect to what 
reasoning processes are performed by an agent, other than the constraints on possible
messages enforced by the commonly agreed solver protocol.
Such privacy of algorithms is described in~\cite{silaghi2002comparison}.
In solutions based on cryptographic protocols the equivalent concept for enforcing privacy of algorithm is introduced in~\cite{cryptoeprint:2005:154,silaghi2005using}, allowing
for the definition of a stronger level of privacy: {\em requested privacy}.

\begin{definition}[requested privacy]
Given secret inputs $\sigma$, the prior knowledge $\Gamma$ of  $t$ colluders and a multi-party computation process $\Pi$ with answer $\alpha$
that can be decomposed in a desired data $\alpha^*$
and an algorithmic dependent unrequested data $\bar\alpha$,
we say that an algorithm $A$ achieves requested $t$-privacy if
the probability
distribution of the secrets 
that an attacker controlling any at most $t$ participants can learn
is conditionally independent on $\Pi$, $A$ and $\bar\alpha$ given requested
data $\alpha^*$
and prior knowledge $\Gamma$.

$$P({\sigma}\mid \alpha,\Gamma,\Pi,A) = P(\sigma\mid \alpha^*,\Gamma)$$
\end{definition}

This concept of requested privacy is the strongest known relevant privacy concept and can be 
relaxed for computational purposes to:

\begin{definition}[non-uniform requested privacy]
Given secret inputs $\sigma$, the prior knowledge $\Gamma$ of  $t$ colluders and a multi-party computation process $\Pi$ with answer $\alpha$ 
that can be decomposed in a desired data $\alpha^*$
and an algorithmic dependent unrequested data $\bar\alpha$,
we say that an algorithm $A$ achieves non-uniform requested $t$-privacy if
for any secret $\bar\sigma \in \sigma$ that is not deterministically revealed given requested data $\alpha^∗$ and prior knowledge $\Gamma$, it is also not deterministically revealed given $\Pi$, $A$, and $\bar{\alpha}$,
to any attacker controlling any at most $t$ participants.

$$\forall \bar{\sigma} \in \sigma P(\bar{\sigma}\mid \alpha^*,\Gamma) < 1 \Rightarrow P(\bar{\sigma}\mid \alpha,\Gamma,\Pi,A) < 1$$

\end{definition}

\item {\bf Privacy of Constraint Topology}

It was shown in~\cite{silaghi2004meeting,silaghi2006secure}
that erasing topological information information
by combining all constraints into a big $n$-ary 
function maximizes privacy. Further, the protection of topological information can be a secret in itself, with application to protecting trade secrets, as in~\cite{silaghi2001generalized,silaghi2002self}.

\item {\bf Agent Identity Privacy}
Hiding the real identity of the agents participating in a DCOP is a concern discussed in~\cite{leaute2013protecting}.
This refers to the general concept of anonymity, where software agents cannot be linked to human owners.
\end{itemize}

\section{Steganographic DCOPs}\label{sec:stega}
Besides the previous types of privacy, we raise the next one:
\begin{itemize}
\item {\bf Privacy of Existence of Secrets}
In an observation we introduce here,
note that for many of the studied DCOP frameworks, the agents
that participate in the computation may need to suggest or admit
that there exist secrets involved in the computation,
without which the additional costs of the setup and
execution would not be warranted.

But admitting that secrets may be involved is sometimes
a significant privacy loss in itself, for example when agents want to claim openness for political reasons.
This concept is introduced in this paper, with the Steganographic DCOP.
\end{itemize}


Steganography is the area of cryptography concerned with
hiding of the existence of secret information.
While the oldest example of steganography dates from Histaeus' 499BC message tattooed under the hair of a slave's head, as described by Herodotus, the best known modern techniques range
from invisible ink to hiding of data in poetry and images.

Steganography is more appropriate then cryptography for 
the case of secrets with social impact, as is often the
case with distributed constraint optimization~(DCOP).
With social impacts, an agent may lose status simply by 
stating that it has secrets, and trying to safeguard
them. Politicians (and not only them) try to claim
openness and lack of secrets. 

In our experience, the need to protect existence of secrets proved to be a key impediment to the large scale adoption of cryptographic DCOP solvers, since many potential
users find it difficult to publicly admit a need, or to
call for secret problem solving.

In this work we identify the steganographic potential of traditional DCOP solvers where the technical distribution of the problem can be claimed as the unique cause of avoiding decentralization and straightforward
full revelation, allowing to hide privacy needs.

The basic framework recommended for steganographic DCOPs,
rephrasing variable ownership requirements as unary constraint, is:

\begin{definition}[StegDomDCOP]
\label{def:StegDomDCOP}
A Steganographic Domain Distributed Constraint Optimization Problem (StegDomDCOP) consists of $n$ agents $A=\{a_1,...,a_n\}$ and $n$ variables $V = \{x_1,x_2,...x_n\}$. 
Each agent $a_i$ has a unary constraint on $x_i$ and its domain $D_i$.
Agents know weighed constraints on subsets of these variables, specifying costs and rewards induced by assignments to concerned variables.
Each agent can also identify, for each cost it associates with a total assignment, a secret reward for not revealing it.

The goal is to choose values for variables such that an objective function is minimized or maximized, while each agent's participation is rational, performing only acts with positive expected sum of costs and rewards. The objective function described is addition over costs, but can be any associative, commutative,
monotonic aggregation operator defined over a totally ordered set of valuations, with
minimum and maximum elements.
\end{definition}

The requirement of having the number of variables equal the number of agents can be dropped, together with the
requirement that each agent has a unary constraint on
a distinct variable, obtaining an equivalent framework that maps easier to certain problems, like to the problems in~\cite{silaghi2000asynchronous}:

\begin{definition}[StegCosDCOP]
\label{def:StegCosDCOP}
A Steganographic Cost Distributed Constraint Optimization Problem (StegCosDCOP) consists of $m$ agents $A=\{a_1,...,a_m\}$ and $n$ variables $V = \{x_1,x_2,...x_n\}$.  Agents know weighed constraints on subsets of these variables, specifying costs and rewards induced by assignments to concerned variables.
Each agent can also identify, for each cost it associates with a total assignment, a secret reward for not revealing it.

The goal is to choose values for variables such that an objective function is minimized or maximized, while each agent's participation is rational, performing only acts with positive expected sum of costs and rewards. The objective function described is addition over costs, but can be any associative, commutative,
monotonic aggregation operator defined over a totally ordered set of valuations, with
minimum and maximum elements.
\end{definition}

The fact that the last two frameworks are equivalent in expression power is guaranteed by the theory of primal dual CSP conversions~\cite{bacchus1998conversion}. However, modeling strategies
and solvers may perform significantly better with one rather than the other approach, since the transformation between representations can be laborious.

The StegDCOP framework has the advantage that its agent problems 
can be  masqueraded as instances of any common DCOP
framework, and can be used in asynchronous search protocols to protect privacy without
making other agents suspicious, as in~\cite{silaghi2002comparison}.

The framework enables the support of privacy of constraints and privacy of domains in a similar way to UDCOP and DCSPs.

\section{Frameworks Disambiguation}\label{sec:frameworks}

Distributed Constraint Optimization Frameworks can be classified along the following independent dimensions:

\begin{itemize}
\item {\bf $X_1$}
 semiring/value system: {\tt Boolean, Fuzzy, Weighted, Probabilistic}
\item {\bf $X_2$}
 structure of the problem: {\tt Static, Dynamic\_Local (D\_L), Dynamic\_Topology (D\_T)}
\item {\bf $X_3$}
 whether domains are known only by some agents, by everyone  with variable ownership (enabling extensional constraint representation), or with distribution of constraints. Combinations are possible. The corresponding distribution reasons are named: {\tt Domains, Variables, Costs, Domains\_Costs (D\_C), ...}
\item {\bf $X_4$}
privacy of decision specification, for opening all assignments in the adopted solution to everyone, or only to specified agents:  {\tt Open, Closed}
\item {\bf $X_5$}
 privacy management, where participants claim no privacy and do not secretly quantify privacy concerns, where participants claim privacy and secretly quantify its loss, where participants claim no privacy but secretly quantify privacy concerns, and where secrecy is admitted and enforced with cryptography: {\tt Public, Quantified, Steganographic, Cryptographic}\footnote{See~\cite{yokoo2002secure,silaghi2003arithmetic,silaghi2004distributed,grinshpoun2019privacy,grinshpoun2016p}}
\item {\bf $X_6$}
optimization function with utilitarian summation of constraints, egalitarian Leximin, or egalitarian Theil index\footnote{See~\cite{matsui2018leximin,netzer2011social}}: {\tt Utilitarian, Leximin, Theil}
\end{itemize}

A nomenclature can be designed to easily differentiate between the different types of frameworks, or to enable studies of framework relations and hierarchy.

To maximize compatibility with past practice and the logic of the framework design, the proposed name structure is:
\begin{equation}
X_6X_5X_4DisX_3X_2X_1COP \label{eq:name}
\end{equation}

where each of the names can be in camel case on any unambiguous prefix of the above properties, to enable disambiguation and easy extensions. 
The new nomenclature~(\ref{eq:name}) is long and shorter versions can be obtained 
by selecting as defaults: $X_6$={\tt Utilitarian}, $X_5$={\tt Public}, $X_4$={\tt Open},
$X_3$={\tt Domain}, $X_2$={\tt Static}, $X_1$={\tt Weighted}.
Each default parameter can be skipped if all other parameters it separates from the base string ``Dis'' are at their default value. 
Further, the BooleanCOP can also be historically named as CSP.

With these conventions, for example:
\begin{itemize}
\item {\bf DCOP.}
The common DCOP in Definition~\ref{def:DCOP}~\cite{modi2002asynchronous} becomes {\tt Utilitarian}-{\tt Public}-{\tt Open}-{\tt Dis}-{\tt Domains}-{\tt Static}-{\tt Weighted}-{\tt COP}
or shorter {\tt UPODisDSWCOP}
With the use of defaults, the acronym becomes {\bf DisCOP}.

\item {\bf UDCOP.}
UDCOPs~\cite{savaux2017utilitarian} with full camel case specifiers becomes
{\tt Utilitarian}-{\tt Quantified}-{\tt Open}-{\tt Dis}-{\tt Domains}-{\tt Static}-{\tt Weighted}-{\tt COP} or shorter, {\tt UQODisDSWCOP}.
With the use of defaults, the acronym becomes {\bf QODisCOP}.

\item 
The new StegDomDCOP framework in Definition~\ref{def:StegDomDCOP} becomes
{\tt Utilitarian}-{\tt Steganographic}-{\tt Open}-{\tt Dis}-{\tt Domains}-{\tt Static}-{\tt Weighted}-{\tt COP}. With the use of defaults, the acronym becomes {\bf SODisCOP}.

\item 
The new StegCosDCOP framework in Definition~\ref{def:StegCosDCOP} becomes
{\tt Utilitarian}-{\tt Steganographic}-{\tt Open}-{\tt Dis}-{\tt Costs}-{\tt Static}-{\tt Weighted}-{\tt COP}. With the use of defaults, the acronym becomes {\bf SODisCCOP}.

\item {\bf DisWCSP.}
The DisWCSP in Definition~\ref{def:DisWCSP}~\cite{silaghi2005using} becomes
{\tt Utilitarian}-{\tt Cryptographic}-{\tt Closed}-{\tt Dis}-{\tt Costs}-{\tt Static}-{\tt Weighted}-{\tt COP} or {\tt UCCDisCSWCOP}.
With the use of defaults, the acronym becomes {\bf CCDisCCOP}.

\item {\bf DCSP.}
With this proposal, the original DCSP in Definition~\ref{def:DCSP}~\cite{yokoo1992distributed} is denoted
{\tt Utilitarian}-{\tt Public}-{\tt Open}-{\tt Dis}-{\tt Domains}-{\tt Static}-{\tt Boolean}-{\tt COP} or {\tt UPODisDSBCOP}.
With the use of defaults, the acronym becomes {\bf DisCSP}, as that acronym was used in several previous 
works~\cite{silaghi2001abt}.

\item {\bf DisCSP}
With the use of defaults, the DisCSPs with private constraints from Definition~\ref{def:DisCSP}~\cite{silaghi2005asynchronous} is denoted
{\tt Utilitarian}-{\tt Public}-{\tt Open}-{\tt Dis}-{\tt Domains}\_{\tt Costs}-{\tt Static}-{\tt Boolean}-{\tt COP} or {\tt UPODisD\_CSBCOP} and would be denoted shortly {\bf DisD\_CCSP}.

\end{itemize}

We note that the term ``{\tt Open}'' that we propose for specifying the publication
of the final decisions, has been used in the past as synonym for dynamism~\cite{silaghi2002openness}.
The term ``{\tt Utilitarian}'' DCOP has been used for specifying explicit representation of Privacy,
but also to express utilitarian maximization of social welfare~\cite{walras1896elements}.
However, the possible acronyms in the new nomenclature will not conflict with prior notations (that did
not already have conflict, like DCSP).

\section{Conclusion}\label{sec:conclusions}
We have identified a new type of privacy relevant to Distributed Constraint Optimization, 
namely {\em privacy of existence of secrets}, leading to the formal introduction of
the {\em Steganographic DCOPs}. 
The new framework allows for disambiguate between problems where privacy is public but quantified and when privacy requirements are hidden (steganography).
We review existing 
frameworks offering privacy in constraint optimization to position the new
type of privacy in the literature, and propose a new nomenclature system that,
if adopted, can help disambiguate all privacy frameworks found in use.
Further, the obtained classification can be used to identify gaps in the literature as well as valuable new frameworks to study.

\bibliographystyle{aaai}
\bibliography{dcops}
\end{document}